# Retro Drug Design: From Target Properties to Molecular Structures

Yuhong Wang*, Sam Michael, Ruili Huang, Jinghua Zhao, Katlin Recabo, Danielle Bougie, Qiang Shu, Paul Shinn, Hongmao Sun*

National Center for Advancing Translational Sciences (NCATS)
9800 Medical Center Drive, Rockville, MD 20850

*Contact information for the corresponding authors:

| | |
|---|---|
| Yuhong Wang | Hongmao Sun, PhD |
| NCATS/NIH | NCATS/NIH |
| 9800 Medical Center Dr. | 9800 Medical Center Dr. |
| Rockville, MD 20854 | Rockville, MD 20854 |
| Phone: 301-480-9855 | Phone: 301-480-9839 |
| e-mail: yuhong.wang@nih.gov | e-mail: hongmao.sun@nih.gov |

**Abstract**: To generate drug molecules of desired properties with computational methods is the holy grail in pharmaceutical research. Here we describe an AI strategy, retro drug design, or RDD, to generate novel small molecule drugs from scratch to meet predefined requirements, including but not limited to biological activity against a drug target, and optimal range of physicochemical and ADMET properties. Traditional predictive models were first trained over experimental data for the target properties, using an atom typing based molecular descriptor system, ATP. Monte Carlo sampling algorithm was then utilized to find the solutions in the ATP space defined by the target properties, and the deep learning model of Seq2Seq was employed to decode molecular structures from the solutions. To test feasibility of the algorithm, we challenged RDD to generate novel drugs that can activate μ opioid receptor (MOR) and penetrate blood brain barrier (BBB). Starting from vectors of random numbers, RDD generated 180,000 chemical structures, of which 78% were chemically valid. About 42,000 (31%) of the valid structures fell into the property space defined by MOR activity and BBB permeability. Out of the 42,000 structures, only 267 chemicals were commercially available, indicating a high extent of novelty of the AI-generated compounds. We purchased and assayed 96 compounds, and 25 of which were found to be MOR agonists. These compounds also have excellent BBB scores. The results presented in this paper illustrate that RDD has potential to revolutionize the current drug discovery process and create novel structures with multiple desired properties, including biological functions and ADMET properties. Availability of an AI-enabled fast track in drug discovery is essential to cope with emergent public health threat, such as pandemic of COVID-19.

**Introduction**

The primary goal of modern drug discovery is to identify molecules of therapeutic benefits. A successful drug molecule usually shares two features: 1. It modulates the biological function of its therapeutic target(s) selectively with sufficient binding affinity; 2. It has a balanced ADMET (absorption, distribution, metabolism, excretion, and toxicity) profile, such that it reaches its target(s) unchanged and with sufficient quantity.  Traditionally, a drug discovery project starts with screening a compound library against a proposed protein target, followed by an optimization process to fix existing issues associated with original hit compounds, such as potency, selectivity, PK, and etc. This traditional drug discovery process requires tremendous input of resources and time.  Computational generation of high-quality drug candidates with desired properties, a long-sought goal of pharmaceutical research, will not only reduce the unprecedented cost of bringing a drug to market dramatically (Paul, Mytelka et al. 2010, Hughes, Rees et al. 2011, DiMasi, Grabowski et al. 2016), it will also dramatically speed up the whole process. Accelerated drug development is of paramount importance for public health threats of pandemics such as COVID-19 (Administration 2020, Grobler, Anderson et al. 2020)

The deep learning (DL) technology (LeCun, Bengio et al. 2015, Schmidhuber 2015) brings hope to a new era of drug discovery and development and have the potential of substantially improving or even revolutionizing the current drug discovery paradigm. DL utilizes multiple layers of neurons to model high-level abstractions, complex and non-linear relation in data, and has outperformed humans in many fields including image processing, text and voice recognition, protein structure prediction and GO game; yet this potential in drug discovery remains to be fulfilled.

Various machine learning and deep learning algorithms have been proposed over the past decade for generation of novel molecules with therapeutic benefits.  Kadurin et al (Kadurin, Nikolenko et al. 2017), Blascheke et al (Blaschke, Olivecrona et al. 2018), and Lim et al  (Lim, Ryu et al. 2018) used autoencoder, variational autoencoder and adversarial autoencoder to identify and generate new molecular fingerprints with predefined properties.  Bjerrum and Threlfall (Esben jannik Bjerrum 2017), Cherti et al (Medhdi Cherti 2017),  and Segler et al (Segler, Kogej et al. 2018) utilized recurrent neural network, in particular, the long short-term memory (LSTM) model  (Hochreiter and Schmidhuber 1997) to generate novel molecular structures with certain target properties.

The autoencoder and RNN models are quite limited for generating novel molecules of preferred properties. These models are not designed to assess, or optimize, the properties of the generated molecules. Furthermore, quality of deep learning models is largely determined by data quality and quantity on which they are based, and unfortunately the sample size and data quality of available experimental drug discovery data are usually insufficient for deep learning methods.

To address the limitations of autoencoder and RNN models, various reinforcement learning (RL) (RS Sutton 1998) and generative adversarial networks (GAN) (Ian Goodfellow 2014) have been proposed and implemented for sequence generation. These models typically consist of sequence generation model, RL, and GAN. The RL and GAN models are used to optimize and move the generated molecules towards the target properties.

Olivecrona et al (Olivecrona, Blaschke et al. 2017) introduced a method to tune a sequence-based generative model for de novo molecular design. Sanchez-Lengeling et al (Benjamin Sanchez-Lengeling 2017) presented ORGANIC, a framework based on both GAN and RL, which is capable of producing a distribution over molecular space that matches with a certain set of desirable metrics. Popova et al (Popova, Isayev et al. 2018) devised and implemented a novel computational strategy for de novo design of molecules with desired properties. As a typical strategy, it includes a generative model that produces chemically valid SMILES string, predictive models that forecast the desired properties of the de novo-generated compounds, and a reinforcement learning module that tips the generated structures to have the desired properties. Putin et al (Putin, Asadulaev et al. 2018) reported a deep neural network, ATNC or Adversarial Threshold Neural Computer, for the de novo design of novel small-molecule organic structures with druglikeness properties. Zhou et al (Zhou, Kearnes et al. 2019) presented a framework, called Molecule Deep Q-Networks (MolDQN), for molecule optimization by combining domain knowledge of chemistry and state-of-the-art reinforcement learning techniques (double Q-learning and randomized value functions). One advantage of MolDQN is that it can produce structures of 100% chemical validity. Zhavoronkov (Zhavoronkov, Ivanenkov et al. 2019) developed a deep generative model, generative tensorial reinforcement learning (GENTRL), for de novo small-molecule design, and GENTRL produced several compounds, which were active in biochemical assays and cell-based assays. Ikebata et al

(Ikebata, Hongo et al. 2017) used Bayesian model to identify promising hypothetical molecules with a predefined set of desired properties.

Most of the efforts in de novo molecular design are based upon deep neural networks, in particular, RNN, GAN, and RL. While these DL methods demonstrated the good potential in drug discovery, they are not ready for prime time in typical drug discovery projects. First, deep learning models require a very large number of good quality samples. Unfortunately, available experimental data for drug discovery are limited in both quality and quantity. Second, current methods are not efficient for sampling molecular structural space. To generate new valid chemical structures, a method can only perform local, small, and slow perturbations on representations like SMILES string (Daylight) or graph; such sampling method is apparently not sufficient considering the vast possible chemical and structural space. This may partially explain why GENTRL took 21 days to produce several active compounds. Sampling regardless chemical structure validity will lead to very small percentage of produced structures to be valid. For example, only 7% of the generated structures by ORGANIC is valid. Third, RL algorithms tend to have difficulties in achieving a good balance of exploration and exploitation, making long term credit assignment, and being unstable likely due to moving target. Fourth, with RL, it is theoretically possible to optimize multiple target properties, but in practice, current implementations use either one target property or weighted sum of multiple target properties.

Here we propose a new strategy called Retro Drug Design (RDD). Unlike the current forward approaches as described above, RDD starts from multiple desired target properties, works backwards, and then generates the "qualified" compound structures.

RDD is based upon following rationales and considerations. First, although the amount of available discovery data of small molecules for drug targets and ADMET properties is insufficient for deep learning models, it is sufficient for traditional, or shallow, machine learning models. Second, over the past decades, we developed a generic fingerprint called ATP of 269 descriptors (Sun 2015). The same ATP descriptors have achieved outstanding performance in traditional machine learning prediction models of all the physicochemical and ADMET properties, accessible to us. ATP is originally designed to have good correspondence with SMILES; in other words, one SMILES produces one ATP, and one ATP corresponds to as few SMILES as possible. Third, many millions of valid compound structures are readily available, and they provide a good coverage of valid chemical space of small molecules.

To test our algorithm, we generated 180,000 ligands for µ opioid receptor (MOR) systems (Al-Hasani and Bruchas 2011), selected and tested 96 in cAMP assay in hMOR-CHO cells. Opioids are the most widely used and effective analgesics for the treatment of pain and related disorders, most commonly used opioids for pain management act on MOR, and the cAMP assay is mature and reliable.

**Methods**

**Molecular representation**

An optimized atom-type-based molecular descriptor system, or ATP, consisting of 221 atom types and 48 correction factors was employed to represent small molecules. The details of the molecular descriptors have been elaborated elsewhere (Sun 2015). Atom types are assigned according to the properties of an atom and its chemical environment. An atom type casting tree was designed to assign atom types, based on whether the atom is aromatic, whether the atom is in a ring, whether the atom is next to different functional groups, etc. This original tree, largely based on a medicinal chemist's intuition, was subject to a recursive optimization cycles in terms of where to further split the tree, where to stop splitting, and where to combine the branches, in order to make the best prediction of *logP* values in the Starlist dataset containing about 11,000 structurally diverse compounds.(Sun 2015) An atom in a molecule is like a piece of puzzle chip in a puzzle, which has its unique edge. When a set of puzzle chips are provided, the puzzle can be solved unambiguously, on the basis of unique shape of each piece.

The optimized tree output 221 atom types, featuring 88 different carbon types, 7 hydrogen types, 58 nitrogen types, 31 oxygen types, 8 halide types, 23 sulfur types, and 6 phosphorus types. (Sun 2015) Forty-eight correction factors are appended to catch a number of whole molecule features, such as the molecular globularity, molecular rigidity, lipophilicity, and etc.(Sun 2015) In total, a series of 269 numerical values comprise the final set of the atom type molecular descriptors.

Using ATP representation, we could design molecules of predefined properties by sampling in the ATP space of 269 dimensions.

**General RDD workflow**

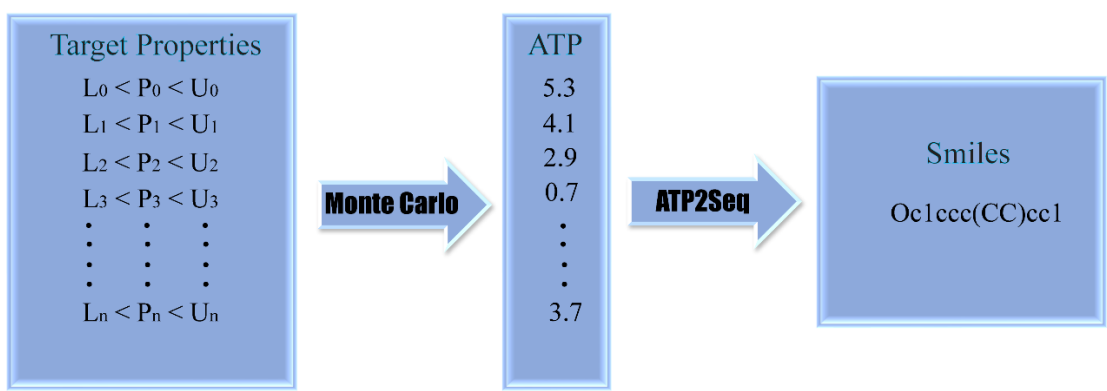

Figure 1. The flowchart for retro drug design.

The workflow of RDD is illustrated in Figure 1. RDD starts from a list of target properties and their preferred ranges, such as logP between 2 and 5, and the number of properties is only limited by available computing resources. The properties can be physicochemical properties, such as molecular size, solubility, and biological properties, such as hERG (Wikipedia) activity. In this study, all properties are computed from the same molecular representation system, ATP, through an evaluator, which could be a simple mathematical function, a traditional machine learning model, or a deep learning neural network model. The property range is defined by lower and upper boundaries of $L_0$-$L_n$ and $U_0$-$U_n$ (Figure 1).

With the input list of target properties and their desired ranges, RDD uses a number of evaluators of $E_0$-$E_n$ (Figure 1) and a Monte Carlo (MC) sampling algorithm to find solutions in the ATP space that have the desired ranges according to the corresponding evaluators. In order to sample in the more likely valid region of the ATP space, we calculated the mean ($ATP\_Mean$) and the standard deviation ($ATP\_Std$) from the ATP for 906,727 unique molecular structures in NCATS's compound collection.

The MC algorithm consists of the following steps:

1. Start from a randomly initialized ATP, $atp$ according to Equation 1

$$atp[i] = ATP\_Mean[i] + gr * ATP\_Std[i] \qquad (1)$$

   Where $gr$ is a Gaussian random number generator with a mean of 0.0 and standard deviation of 1.0. Apply the evaluators and compute the output score $S_i$. Then calculate the initial cost according to Equation 2.

$$C = \sum_{i=0}^{n} w_i(L_i - S_i)_{S_i<L_i} + w_i(S_i - U_i)_{S_i>U_i} \qquad (2)$$

   Where n is the number of properties or evaluators, $w_i$ is the weighting factor for the $i$th evaluator, $L_i$ and $U_i$ are the lower and upper boundaries of the $i$th property.

2. Randomly pick 8 elements (optional) from previous $atp$, and randomly perturb each by adding 0.5*($r$ - 0.5) * $ATP\_Std$[i] to propose a new $atp$. $r$ is a uniform random number generator between 0.0 and 1.0.

3. Recalculate the cost of the current $atp$. If the current cost is < 0.01 (optional), stop the sampling process and output the solution. If the current cost is smaller than the previous one, accept the perturbation; otherwise reject the perturbation. Go back to step 2.

4. If no solution is found after 50 steps, go back to step 1.

5. If no solution is found after 40,000 steps, stop and terminate.

With the solution of *atp* in the ATP space from the Monte Carlo algorithm, next, RDD uses a deep learning model, called ATP2SMI, to map *atp* to molecular structures as represented by SMILES. ATP2SMI is adapted from the widely used seq2seq model (Google). A typical seq2seq model consists of an encoder and a decoder; the encoder transforms the sequence input to a vector of latent variables, which is then transformed to the output sequence by the decoder. In ATP2SMI, we removed the encoder, used ATP as the vector of latent variables, and then used a decoder to transform ATP to output sequence-SMILES (Figure 2).

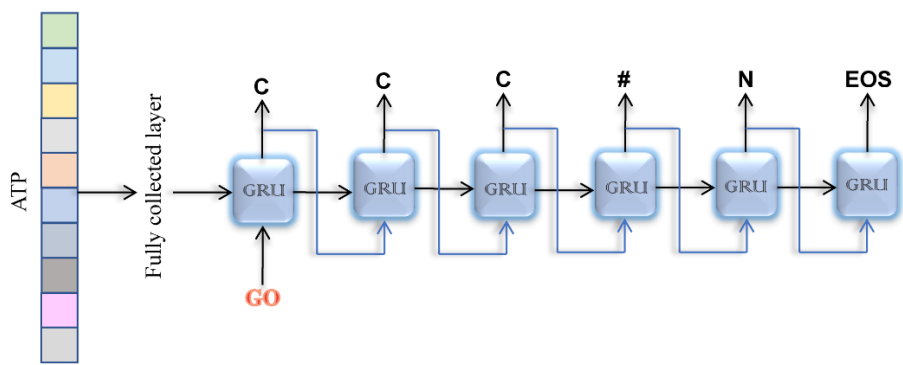

Figure 2. ATP2SMI network. The vector of ATP is fed to a fully connected layer, FC, and the output is fed to decoder, a recurrent neural network of GRU cells. Each GRU cell outputs a letter of a SMILES.

SMILES uses a string of letters to represent a molecular structure. In this study, we ignored chirality in a molecular structure, and used a vocabulary of only 40 letters or words -- A, T, E, U, = # % ( ) [ ] / \ 0 1 2 3 4 5 6 7 8 9 B C N O P S F I c n o p s - + Br Cl -- to encode a SMILES. A, T, E, U are added for convenience; A is for padding, T for start, E for stop, and U for any letters not included in this vocabulary list. Each element in this vocabulary list is encoded by a one-hot vector. We ignored chirality in a molecular structure, mainly for simplification purpose, and plan to include chirality in the future studies.

In the ATP2SMI, an *atp* is fed to a fully connected layer, FC, and the output is fed to decoder, a recurrent neural network of GRU cells. The number of units in the FC is the same as that in the GRU cell, and we tried four different numbers of units, 1024, 1280, 1536 and 2048.

Sparse categorical cross entropy is used as the loss or cost function. Backpropagation is used for training the network (David E. Rumelhart 1986). Optimization of the loss function is carried out by mini-batch of a size 128 and the ADAM optimizer (Diederik P. Kingma 2017), which is implemented as tf.train.AdamOptimizer in the Tensorflow library (Tensorflow). For the ADAM optimizer, a learning rate of 0.001 produced a satisfactory result.

The model training process was monitored by two metrics: cost function and accuracy on both the training and validation datasets. The model with the best accuracy on the validation set is saved and applied to the test dataset to collect chemistry specific benchmarks. In this study, we compute three chemistry specific metrics. The first is the percentage of the generated structures that are chemically valid. A structure is considered valid if it is successfully parsed by ChemAxon's molecular parser (Chemaxon). The second is the percentage of the valid structures that are identical to the ground truth. The third is the percentage of the structures that have a Tanimoto score > 0.95.

**SVM models as evaluators of ligand binding activity for MOR and blood brain barrier (BBB) permeability**

In order to recognize ligands for MOR, a SVM model (Noble 2006) was constructed using the primary screening results.

Among the 1707 preprocessed samples, 256 were assigned active and 1451 inactive. Apparently, this dataset was highly skewed. To address this data imbalance problem, we adopted the bootstrapping and Jackknifing method (Sun, Huang et al. 2017). Briefly, the 1707 samples were randomly split into training dataset (80%) and test dataset (20%). The negative samples in the training dataset were further randomly split into 5 subsets, which were then combined with the positive samples to produce 5 sub training datasets. Five SVM models were trained using the software package of LIB-SVM (C-C Change 2001). The parameterization of the penalty for misclassification, *C*, and the non-linearity parameter in the kernel function of a Gaussian Radial Basis Function (RBF), $\gamma$, was accomplished on a grid-based search to minimize the mean standard error (MSE) of 5-fold cross-validation (CV) on the training data. The consensus prediction results on the test dataset were used for benchmarking the accuracy and AUC-ROC (Hughes, Rees et al. 2011). The whole process was repeated for 10 times. The confidence intervals of the accuracy and AUC-ROC were 0.812 ±0.0063 (95%) and 0.863 ±0.0167 (95%), respectively.

BBB classification model was built with the same ATP descriptors, based on a curated data set of 1964 compounds (Martins, Teixeira et al. 2012). The SVM model trained with 80% randomly selected compounds achieved a high predictivity on the 20% test set, with AUC-ROC of 0.92 and accuracy of 0.87.

**cAMP assay in hMOR-CHO cells**

3-isobutyl methyl xanthine (IBMX), NKH 477 and naloxone hydrochloride were purchased from Sigma–Aldrich (St. Louis, MO). DAMGO was purchased from Abcam Inc. (Cambridge, MA). cAMP-Gi assay kit was from Cisbio (Bedford, MA). Human recombinant μ opioid receptor (hMOR-CHO-K1) stably express mu cell line was purchased from Multispan, Inc. (Hayward, CA). All the cell culture reagents were obtained from Invitrogen (Life Technologies, Madison, WI). hMOR-CHO-K1 cells were cultured in DMEM/F12 medium supplement with 10 % fetal bovine serum, 100 U/mL penicillin, 100 μg/mL streptomycin and 10 μg/mL puromycin at 37°C under a humidified atmosphere and 5% $CO_2$.

hMOR-CHO cells were re-suspended in culture medium and dispensed at 2,000 cells/3μL/well in 1,536-well white plates (Greiner Bio-One North America, Monroe, NC) using a Multidrop Combi (Thermo Fisher Scientific Inc., Waltham, MA). After incubation at 37°C for 18 h, 23 nL of compound dissolved in dimethyl sulfoxide (DMSO) or DMSO only was added to the assay plates via a Wako Pintool station (Wako Automation, San Diego, CA). Following compound addition, 1 μL of IBXM at final concentration of 0.5 mM were transferred to the assay plate by a BioRaptr Flying Reagent Dispenser (FRD) (Beckman Coulter, Brea, CA). The assay plates were then incubated at 37°C for an additional 30 min. And then 2.5 μL of cAMP-d2 and 2.5 μL anti cAMP-Cryptate were added to each well and the assay plates were incubated at room temperature in the dark for 1 h. The fluorescence intensity of the assay plates was measured at 340 nm excitation and 665 and 620 nm emission using an Envision plate reader (Perkin Elmer, Boston, MA). Data was expressed as the ratio of 665nm/620nm emissions. Each test compound was tested in triplicates at 11 concentrations ranging from 0.00078 to 46.08 µM.

**Computer hardware and software**

The computations were performed on a Dell PowerEdge R940xa server with four Intel Xeon Platinum 8160 processors (each with 24 cores), 3TB of RAM and four 16GB NVIDIA Tesla V100 graphic processing unit, installed with Ubuntu 16.04.6 distribution, python 3.5, CUDA driver version 10.0, cuDNN version 7.4, TensorRT 5.1 and TensorFlow 1.13.1. A Java program was written to use JOELib (JOELib) for molecule structure parsing and ATP calculation, a Java program was written to implement the Monte Carlo algorithm, and a python script was written to implement the ATP2SMI model.

**Results and Discussion**

*ATP2SMI*

We used entire 906,727 unique molecules in NCAT's compound collection to train the ATP2SMI model to learn the general principles of the molecular system. This collection is of pharmaceutical interest, consisting of the marketed drugs, drugs that have reached clinical trials, and other bioactive molecules. Each ATP and the corresponding ground truth SMILES form a data sample of input and output. The 906,727 samples were randomly split into a training dataset of 816,424 samples (90%), a validation dataset of 45,306 samples (5%), and a test dataset of 44,998 samples (5%).

The training process took about four days for a GRU cell of 2,048. The cost and the accuracies versus epoch on the training and validation datasets were plotted in Figure 1. The cost dropped dramatically in the first 10 epochs and then continued to decrease slowly. The accuracies on both training and validation datasets also increased rapidly in the first 10 epochs, and then continued to improve slowly.

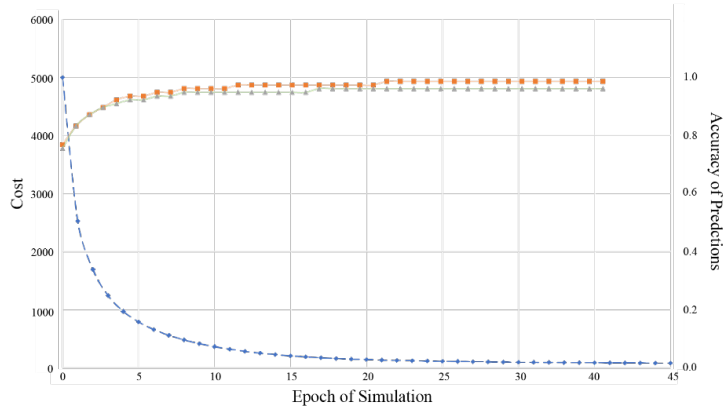

Figure 3. Cost and accuracies vs epoch on the training and validation datasets. The number of the units in both the fully connected layer and the GRU cell is 2048.

We trained 4 ATP2SMI models using GRU cells of 1024, 1280, 1536 and 2048 units, and the accuracies of the optimized models on the test dataset are given in Table 1. The accuracy improves as the unit number of the GRU cell in the decoder increases.

For testing the ATP2SMI model, we used the unit size of 2048. The test dataset has 44,998 samples. Among the generated 44,998 SMILES, 42,053 (93.5%) are chemically valid, and among the 42,053 valid SMILES, 19,716 (46.9%) are identical to the ground truth and 40,762 (96.9%) have a Tanimoto score > 0.95.

When ATP descriptor system was designed, one of the major motivations was to produce a molecular descriptor system that is universal, in other words, the same system can be applied to generate QSAR models for all properties.(Sun 2004) Our following efforts have proved that ATP can provide excellent QSAR models for all the datasets available to us, and most of the QSAR models achieved accuracy comparable to experimental determinations.(Sun 2005, Sun, Veith et al. 2011, Sun 2015, Sun, Huang et al. 2017, Sun, Nguyen et al. 2017, Sun, Shah et al. 2019, Sun, Wang et al. 2020) To achieve a universal molecular descriptor system, the descriptors should well represent the compound and extract its chemical meanings as accurately and adequately as possible. Atom typing meets these requirements. As long as each atom type carries enough information of its surrounding atoms and bonds, ATP descriptors and the corresponding SMILES is interconvertible. Therefore, choosing ATP is critical for the success of RDD.

**Generation of ligands for μ opioid receptor (MOR) with BBB permeability**

Opioids are the most widely used and effective analgesics for the treatment of pain and related disorders, and most commonly used opioids for pain management act on MOR. We chose MOR and the cAMP assay in hMOR-CHO cells to test our algorithm because the cAMP assay is available to us in-house and it is mature and reliable. Furthermore, opioid addiction is a huge challenge to worldwide public health, and this effort to generate potential ligands from RDD for MOR is part of the champaign of the HEAL program (NIH).

MOR is a G-protein-coupled receptor (GPCR) and the target of most opioids, such as morphine. The crystal structures of MOR revealed a vast ligand binding pocket, which could accommodate small molecule agonists and peptide agonists, such as peptide MANGO (Figure 4). This can explain the high hit rate of the primary assay. In this study, we aimed at generating MOR agonists with BBB permeability. Since small-size molecules with a polar surface area less than 60 Å² have a better opportunity to penetrate the BBB by passive diffusion, RDD will be induced to output MOR agonists of small or medium size.

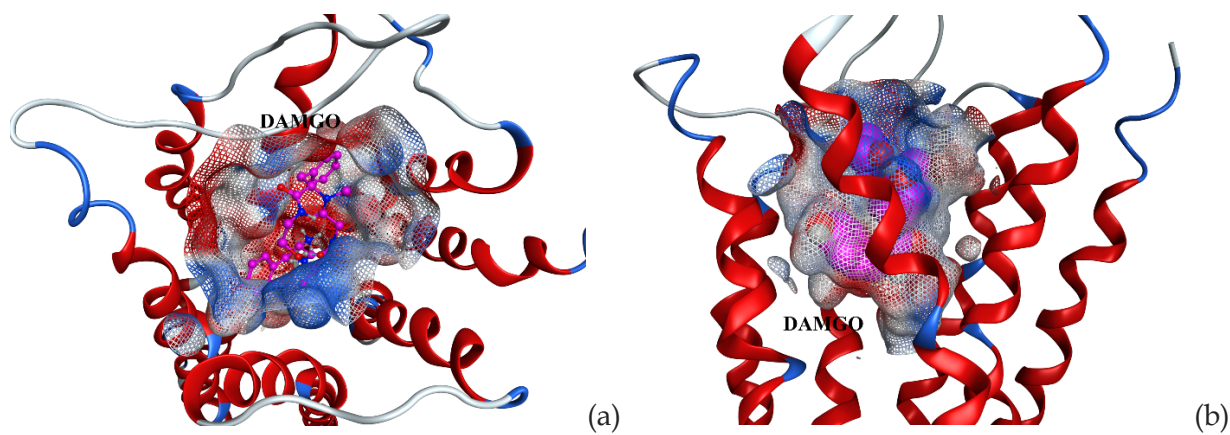

(a)  (b)

Figure 4. Top view (a) and sideview (b) of the ligand binding site of MOR in complex with peptide agonist DAMGO.

In order to test the RDD algorithm, we used three target properties. The first property was the molecular size as measured by the number of heavy atoms. Most of the approved small molecule CNS drugs have 27-37 heavy atoms, so the lower and upper boundaries are set at 27 and 37. The corresponding evaluator was a simple function taking sum of the first 221 descriptors in ATP for non-hydrogen atoms. The second property was the probability of a molecule to modulate MOR, which was evaluated with the SVC model built on internal MOR primary assay results, as described earlier. The lower and upper boundaries were set between 0.5 and 1.0. The third property was the probability of a molecule being permeable to BBB, which was evaluated with the SVC model built on the curated BBB dataset as described earlier. The lower and upper boundaries were also set between 0.5 and 1.0.

The Monte Caro sampling was performed in the ATP space. We generated 180,000 SMILES, and on our workstation, it took about 200 CPU hours. On average, it takes four seconds to find one solution in the ATP space, 5 seconds to generate one valid structure (4/0.78), and 16 seconds to generate one structure with three desired target properties (5/0.31).

Of the 180,000 generated structures, 139,720 SMILES (78%) are chemically valid. Out of the 139,720 valid SMILES, 58,879 (42%) have heavy atom number between 27 and 37, 137,624 (98.5%) and 107,505 (77%) are predicted to be BBB permeable and MOR active, respectively. There are 42,653 (31%) SMILES that meet three requirements.

To generate a valid SMILES from scratch is not trivial. There are numerous underline rules to follow, in order to avoid 5-carbon aromatic rings, or 5-bond carbons. RDD can learn these chemical and structural rules and incorporate them into the process of generating new structures, as indicated by the results. Furthermore, nearly one third of the RDD generated compounds met all three predefined conditions.

Chemical space is vast, like a galaxy. The past decade observed tremendous efforts to expand the coverage of both physical and virtual chemical space.(Hoffmann and Gastreich 2019) Merck

MASSIVE 2018 contains $10^{20}$ virtual compounds, seconded by AZ space of $10^{17}$ capacity. (Hoffmann and Gastreich 2019) Even though computer software and hardware can handle these huge compound libraries, they are ignorable comparing with the size of the estimated druglike chemical space of $10^{60}$. To rebuild the galaxy is not an efficient way for drug hunting, although occasional successes have been achieved.(Lyu, Wang et al. 2019) RDD takes a different approach by locating the relevant constellation instead of exploring the whole galaxy. In this study, three conditions, MOR activity, BBB permeability, and molecular size, defined a subspace, or constellation, and RDD acknowledged the conditions and generated novel structures within the constellation.

**Confirmation of the RDD hits by experiments**

Due to the limited the resources, we can only afford the purchasable chemicals. As a result, we computed the InChi hash keys for the 139,720 structures, searched about 20 millions of small molecular structures in the Sigma-Aldrich catalog, and found about 267 matches. Among the 267 matches, only 96 are available for purchase (Table 2). These 96 compounds were not high ranking ones among the 107,505 structures that were predicted to be MOR active. The highest ranking commercially available hit ranked 2,986 in the 139,720 RDD generated structures (Figure 5). The striking feature observed in this study was the high percentage of the novel structures generated by RDD – only 267 out of nearly 140,000 RDD hits were commercially available. RL and GAN usually start from existing molecules and introduce different level of perturbation to generate new structures, so the new structures are generally close analogues of their templates. RDD started from random numbers, so the algorithm will not be restricted by existing seeds in terms of chemical space. In this sense, RDD is the authentic tool to explore the sub-space relevant to the targets and predefined properties unbiasedly. The advantage of RDD algorithm is obvious, i.e., it avoids searching the whole galaxy, which is inefficient assuming it is possible, for the target, instead, it explored only the small chemical space related to the targets. Another truth revealed by this study is the collection of commercially available compounds is tiny, and the vast chemical space is insufficiently investigated.

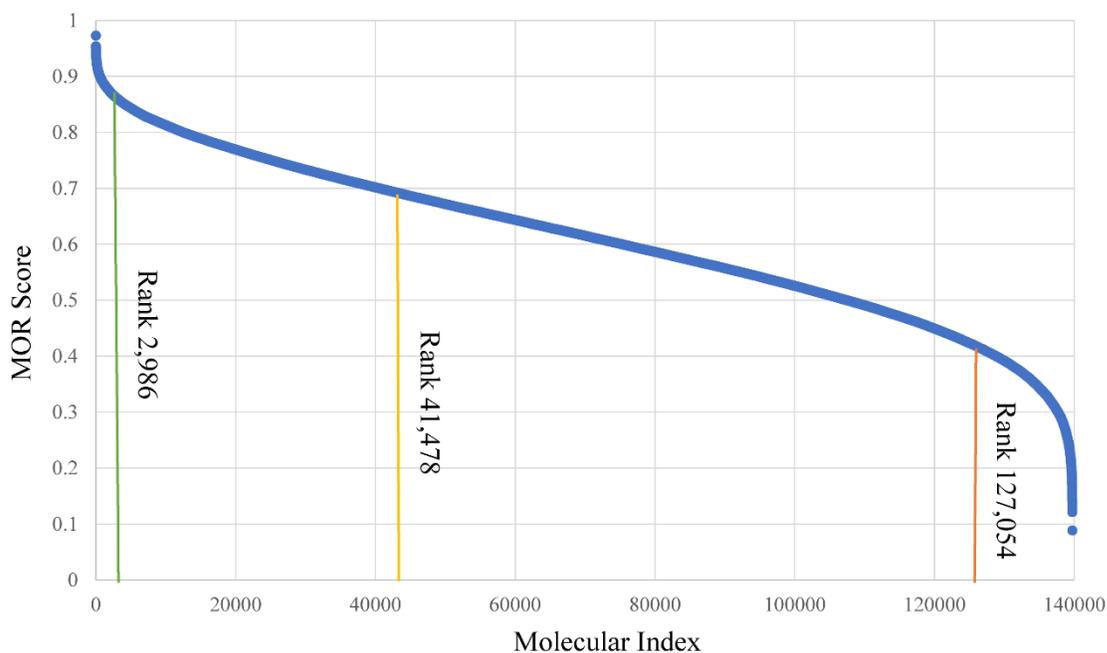

Figure 5. Distribution of MOR scores for 139,720 RDD generated hits. The green, yellow and red lines indicate the highest, median and lowest ranking of MOR score for the commercially available hits.

We purchased these 96 compounds and performed the cAMP assay in hMOR-CHO cells. At NCATS, a sample is defined as positive if its curve class belongs to (1.1, 1.2, 2.1, 2.2) or its efficacy is > 50%. According to this definition, 25 out of 96 were found MOR positive, and the positive hit rate was 26%. Four of the 25 confirmed hits had a curve class of 1.1 (Figure 6), and the $AC_{50}$ ranged between 5.25 and 41.76 ( µM).

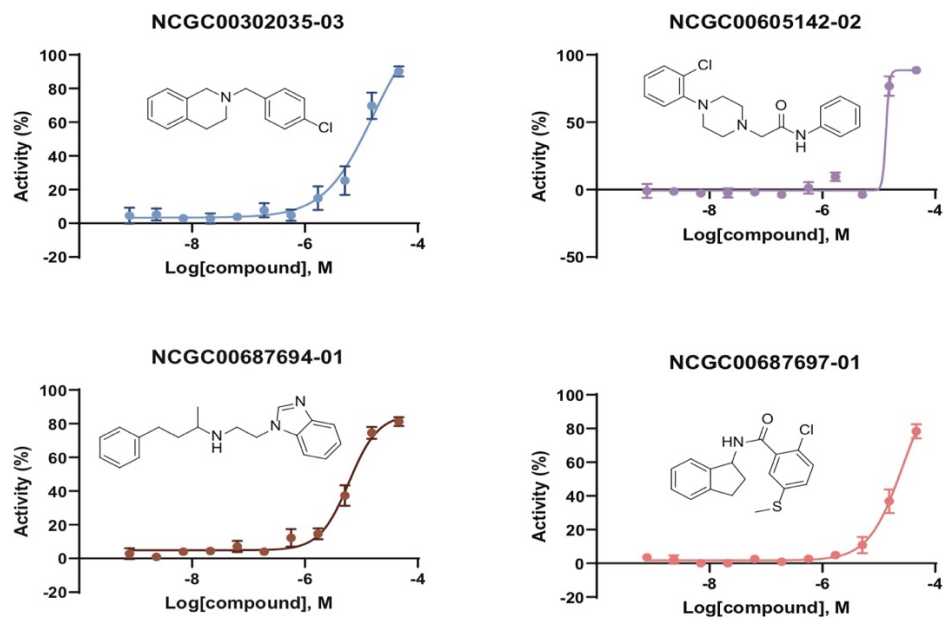

Figure 6. Dose-response curves and chemical structures of the four generated structures with curve class of 1.1.

The 25 generated and active molecules were not structurally similar to the 256 active compounds from the primary assay. Tanimoto similarities between the 25 RDD hits and their closest analogues in the 256 primary assay hits were calculated, and they were between 0.24 and 0.41. Since atom typing was to dismantle a molecule to small pieces, such as atoms and functional groups, and SVM and other ML algorithms extracted the important or discriminant atom types associated with a protein target or a property, then Monte Carlo algorithm served as a search engine to search for the solutions to the multiple conditions in atom type space, dissimilarity of the final solutions resulted from the process of reassemble the atomic pieces back to a molecule.

In summary, we demonstrated that RDD was able to generate novel chemical structures with multiple targeted properties. For MOR, we achieved a hit rate of 26% even though the commercially available compounds did not rank top for activity against MOR. All the 25 generated active compounds are structurally dissimilar from the active compounds from the primary screening. On average, RDD generated one compound with desired target properties in 16 seconds on a single CPU core.

Designed as a platform, RDD supports various evaluators/plugins, and it allows multiple properties to be optimized individually and yet at the same time. Most generative models are theoretically capable of optimizing multiple properties at the same time; in their current implementation, however, they only support using one target property or weighted sum of multiple properties as cost function.

These features bring a number of advantages for RDD. First, to generate new structures with desired ADMET properties, RDD does not need a very large quantity of quality samples. A typical machine learning model such as SVM could be trained over a specific experimental dataset of moderate size, as few as a few hundreds. Second, RDD does not use weighted sum of properties as cost function, and it allows multiple properties to be individually optimized. In this study, we only used three evaluators for target properties. In another unpublished study, we used 10 evaluators for 10 target properties and achieved comparable results. Third, the Monte Carlo sampling algorithm is very stable and fast. Very few processes failed to produce solutions in the ATP space, and it takes about 4 seconds to find one solution. Even with 10 target properties, it just takes about 30 seconds to find a solution.

**Conclusion:**

Different from searching for a star in a vast galaxy, the strategy that the traditional drug discovery takes, RDD locates the star by exploring its neighboring subspace only. In this study, we have demonstrated that the RDD platform is capable of generating highly novel structures from scratch to meet predefined requirements, including MOR activity and BBB permeability. Nearly 180,000 chemical structures were generated from random numbers, among which 78% were chemically valid. About one third of the valid structures fell into the property space defined by MOR activity and BBB permeability. Out of the 42,000 qualified structures, only 267 chemicals were commercially available, indicating a high extent of novelty of the AI-generated compounds. We purchased and assayed 96 compounds, and 25 of which were found to be MOR agonists. These compounds were structurally diverse and had excellent BBB scores. The results presented in this paper illustrate that RDD has potential to greatly improve efficiency of drug discovery process, by creating novel structures with desired biological and ADMET properties. Availability of highly efficient and productive drug discovery platform is essential to handle emergent public health threat, such as pandemic of COVID-19.

**Acknowledgment:**

This research was supported in part by the Intramural/Extramural research program of the NCATS, NIH.

Table 1. Accuracies of the optimized models on the test dataset.

| Epoch | ATP_LV Dimension | Number of units in GRU cell | Accuracy |
|---|---|---|---|
| 120 | 7 | 1024 | 0.885 |
| 40 | 7 | 1280 | 0.885 |
| 30 | 7 | 1536 | 0.885 |
| 70 | 7 | 2048 | 0.897 |
| 120 | 14 | 1024 | 0.923 |
| 40 | 14 | 1280 | 0.923 |
| 110 | 14 | 1536 | 0.936 |
| 30 | 14 | 2048 | 0.936 |
| 250 | 38 | 1024 | 0.962 |
| 80 | 38 | 1280 | 0.962 |
| 50 | 38 | 1536 | 0.962 |
| 120 | 38 | 2048 | 0.974 |

Table 2. SMILES of the 96 compounds purchased from Sigma-Aldrich

FC1=CC=C(C=C1)N2CCN(CC(=O)N3CCC4=C3C=CC=C4)CC2
CCC1=CC=C(OCC(=O)N2CCC3=CC=CC=C23)C=C1
ClC1=CC=C(C=C1)C2=CC3=C(C=CC=C3)N=C2
CC1=CC(OCCN2C=NC3=CC=CC=C23)=CC=C1
COC1=CC=C(C)C=C1NC(=O)CN2CCCC2C3=CC=C(F)C=C3
FC1=CC=C(CN2CCC3=CC=CC=C3C2)C=C1
COC1=CC(=CC=C1)C(=O)NC2=CC(=CC=C2)N3CCCC3
NC1=CC=C(C=C1)C2=NC3=CC=CC=C3S2
COC1=CC=C(C=C1)C2CCCN2CC3=CC=C(F)C=C3
FC1=CC=C(CN2CCN(CC2)C3=CC=CC=C3)C=C1
CN1CCN(CC1)C(=O)C2C3=CC=CC=C3OC4=CC=CC=C24
FC1=CC=C(C=C1)N2CCN(CC2)C(=O)CC3=CC=CC=C3
CSC1=CC=CC=C1NC(=O)CCN2CCC(C2)C3=CC=CC=C3
CCOC1=CC=C2N(CC3=CC=CC=C3)C(C)=C(C(C)=O)C2=C1
ClC1=CC=CC=C1CC(=O)N2CCN(CC2)C3=CC=CC=C3
CSC1=CC=CC=C1C(=O)NCCC2=CC=C3OCCOC3=C2
FC1=CC=CC(CC(=O)N2CCCC2C3=CC=C(Cl)C=C3)=C1
CN1CCC(C2=CC=CC=C2)C3=CC=CC=C13
COC1=CC=CC(=C1)C(=O)N2CCCC3=CC=CC=C23
FC1=CC=CC=C1N2CCN(CC(=O)NCC3=CC=CC=C3)CC2
CN1CCN(CC1)C2=CC=C(C=C2)C(=O)C3=CC=C(F)C=C3
COC1=CC=C(C=C1)C(=O)N2CCCC3=CC=CC=C23
NC1=CC=C(C=C1)N2CCC3=CC=CC=C23
COC1=CC=CC=C1N2CCN(CC2)C(=O)CC3=CC=C(Cl)C=C3
COC1=CC=C(C=C1F)C(=O)NC2=CC=CC=C2N3CCN(C)CC3
NC1=CC=C(CC2=CC=CC=C2)C=C1
CC1=CC=CC(NC2=NC(=CS2)C3=C/C4=CC=CC=C4OC\3=O)=C1
CC(CCC1=CC=CC=C1)NCCN2C=NC3=CC=CC=C23
CN1CCC(CC1)OC(=O)C2C3=CC=CC=C3OC4=CC=CC=C24

```
NC1=C2CCCC2=NC3=CC=CC=C13
CN(CC1=CC=CC=C1)C2CCC3=C2C=CC=C3
CSC1=CC=C(Cl)C(=C1)C(=O)NC2CCC3=CC=CC=C23
CC1=CC=C(OCCC(=O)NC2=CC=CC(=C2)C#N)C=C1
CC1CCN(CC1)C2=CC=CC=C2C#N
CC1=CC=C(OCC(=O)N2CCN(CC2)C3=CC=CC=C3)C=C1
CC1=C(C=CC=C1)N2CCN(CC2)C(=O)C3=CC=C(Cl)C=C3
CN(CC1=CC=C(Br)C=C1)C2CCN(C)CC2
CSC1=CC=CC=C1C(=O)N2CCN(CC2)C3=CC=CC(Cl)=C3
CN(CC1=CC=CC=C1)C(=O)COC2=CC=CC=C2Cl
COC1CCCN(C1)C2=CC=C(NC(=O)C3=CC(F)=CC=C3)C=C2
FC1=CC=C(CN2CCN(CC2)C3=CC=CC(Cl)=C3)C=C1
O=C(CSC1=CC=C2C=CC=CC2=C1)N3CCCC3
FC1=CC=C(C=C1)N2CCN(CC2)C(=O)CC3CCC4=CC=CC=C34
CC1=CC=CC=C1N2CCN(CC3=CC=CC=C3F)CC2
BrC1=CC=C(NC(=O)CN2CCCC3=CC=CC=C3C2)C=C1
CC(C)C(=O)N1CCCN(CC2=CC=CC=C2F)CC1
C(N1CCC2=CC=CC=C12)C3=CC=CC=C3
ClC1=CC=CC=C1C2=CC=CC=C2
ClC1=C(C=CC=C1)N2CCN(CC(=O)NC3=CC=CC=C3)CC2
CN1CCC(CC1)OC2=CC=C(NC(=O)C3=CC=CC=C3F)C=C2
C(N1CCCC(C1)C2=CC=CC=C2)C3=CC=CC=C3
ClC1=CC=C(C=C1)C2=NC(N3CCCC3)=C4C=CC=CC4=N2
FC1=CC=CC(NCC(=O)N2CCC3=CC=CC=C3C2)=C1
CCC(=O)N(C)C1CCCN(C1)C2=CC=CC=C2
COC1=CC=C(CN2CCN(CC2)C3=CC=CC=C3F)C=C1OC
ClC1=CC=C(C=C1)N2CC3=CC=CC=C3C2
CSC1=CC=CC=C1C(=O)N2CCC3=CC=CC=C23
ClC1=CC=C(SCC(=O)NC2CCCC3=CC=CC=C23)C=C1
CC(NC(=O)CC1=CC=CC=C1Cl)C2=CC=CC3=C2C=CC=C3
ClC1=CC=C(C=C1)C(=O)N2CCN(CC2)C3=CC(Cl)=CC=C3
ClC1=CC=C(CN2CCC3=CC=CC=C3C2)C=C1
```

CCCOC1=CC=C(CN2CCCC2C3=CC=C(F)C=C3)C=C1
COC1=CC(=CC=C1)C(=O)NCCC2=CC=C(C=C2)N3CCCC3
FC1=CC=C(CC(=O)NCC2(CCCC2)C3=CC=CC=C3)C=C1
ClC1=CC(=CC=C1)C2=NC3=CC=CC=C3S2
CSC1=CC=C(CC(=O)N2CC(=O)NC3=CC=CC=C23)C=C1
CCOC1=CC=C(CN2CCCC2C3=CC=C(F)C=C3)C=C1
CC(=O)C1=CC=CC=C1NCC(=O)N2CCC3=CC=CC=C23
FC1=CC=CC=C1N2CCN(CC2)C(=O)CCC3=CC=CC=C3
CSC1=CC=CC=C1NC(=O)CCN2CCC3=CC=CC=C3C2
COC1=CC=C(C=C1)C(=O)N2CCN(CC2)C3=CC(Cl)=CC=C3
FC1=CC=C(NC(=S)N2CCC3=CC=CC=C23)C=C1
FC1=CC=CC(NC(=O)CCN2CCC(C2)C3=CC=CC=C3)=C1
COC1=CC=CC=C1NC2=NC(=CS2)C3=CC=C(C)C=C3
O(C1=CC=CC=C1)C2=CC=C(C=C2)C3=CC=CC=C3
COC1=CC=CC(=C1)C(=O)NC2=CC=CC=C2N3CCCC3
CN1CCCN(CC1)C(=O)C2=CC=C(Cl)C=C2
CC1=CC2=C(O)C=C(CSC3=CC=CC=C3)N=C2C=C1
CN(CC(=O)NC1CCCC2=CC=CC=C12)CC3=CC(Cl)=CC=C3
COC1=CC=CC=C1CCNC(=O)CN2CCC3=CC=CC=C23
CC1=CC=CC=C1OCC(=O)N2CCC3=CC=CC=C3C2
ClC1=CC=CC(CN2CCC3=CC=CC=C3C2)=C1
CC1=CC=C(C=C1)N2COC3=CC=C(Br)C=C3C2
O=C(CCC1=CC=CC=C1)N2CCCC3=CC=CC=C23
CSC1=CC=C(CN(C)CC(=O)N2CCC3=CC=CC=C23)C=C1
CC(C)N1CCC(CC1)NC(=O)CC2=CC=C(Cl)C=C2
O=C(CN1CCCC2=CC=CC=C12)NC3=CC=CC=C3
ClC1=CC=C(CCNC(=O)C2(CC2)C3=CC=CC=C3)C=C1
CSC1=CC=CC=C1C(=O)N2CCCC3=CC=CC=C23
ClC1=CC=C(C=C1)N2CCN(CC2)C(=O)C3=CC=C4OCOC4=C3
FC1=CC=C(CCNC(=O)C2(CC2)C3=CC=CC=C3)C=C1
CSC1=CC=C(CN(C)C(=O)C2=CC=C3C=CC=CC3=C2)C=C1
OC1=C(C=NC2=CC(Cl)=CC=C12)C(=O)N3CCCC3

ClC1=CC=CC(CN2N=C(C=CC2=O)C3=CC=CC=C3)=C1

ClC1=CC=C(C=C1)N2CCN(CC3=CC=CC=C3)CC2

CSC1=CC=CC(NC(=O)C2=CN(C)C3=CC=CC=C23)=C1